\documentclass[superscriptaddress,twocolumn]{revtex4}
\usepackage{bbm}
\usepackage{amssymb}
\usepackage[tbtags]{amsmath}
\usepackage{graphicx}
\usepackage{epsfig,graphicx,times}


\begin{document}

\title{Does the Berry phase in a quantum optical system originate from the
rotating wave approximation $?$}
\author{Minghao Wang}
\email{sesiseu@gmail.com}
\affiliation{Institute of Theoretical Physics, Shanxi University, Taiyuan 030006, China}
\affiliation{Quantum Optoelectronics Laboratory, School of Physics Science and
Technology, Southwest Jiaotong University, Chengdu 610031, China}
\author{L. F. Wei}
\affiliation{Quantum Optoelectronics Laboratory, School of Physics Science and
Technology, Southwest Jiaotong University, Chengdu 610031, China}
\affiliation{State Key Laboratory of Optoelectronic Materials and Technologies, School of
Physics and Engineering, Sun Yat-Sen University, Guangzhou 510275, China}
\author{J. Q. Liang}
\affiliation{Institute of Theoretical Physics, Shanxi University, Taiyuan 030006, China}
\date{\today }

\begin{abstract}
The Berry phase (BP) in a quantized  light field demonstrated more than a decade ago
 ( Phys. Rev. Lett. \textbf{89}, 220404) has attracted considerable attentions, since it
  plays an important role in the cavity quantum electrodynamics.
 However, it is argued in a recent paper ( Phys. Rev. Lett. \textbf{108}, 033601)
 that such a BP is just due to
the rotating wave approximation (RWA) and the relevant BP should
vanish beyond this approximation. Based on a consistent analysis
we conclude in this letter that the BP in a generic Rabi model actually exists, no matter
 whether the RWA is applied. The existence of BP is also generalized to a
three-level atom in the quantized cavity field.\vspace{0.3cm}

PACS number(s): 42.50.Pq, %Cavity quantum electrodynamics; micromasers
03.65.Vf, %Phases: geometric; dynamic or topological
42.50.Ct.
%Quantum description of interaction of light and matter; related experiments
\end{abstract}

\maketitle

\section{Introduction}

Thirty years ago Berry discovered that, when a quantum system varies slowly
around a closed loop in a certain parameter space the eigenstate of
Hamiltonian will acquire a geometrical phase in addition to the usual
dynamic phase \cite{Berry}. Shortly after, this phase, called usually Berry
phase (BP), was generalized to various versions~\cite%
{AA,ExBerry1,ExBerry2,ExBerry3}. Now, it is found that geometric phases are
related to many physical problems such as Aharonov-Bohm effect, quantum Hall
effect, and Born-Oppenheimer approximation, etc.~\cite{book}. Hopefully,
geometric phases play an important role in fault-tolerant quantum computing~%
\cite{App1,App2}.

The most BPs investigated previously are in the semiclassical context,
namely, the quantum systems are driven by a classical field. In 2002,
Fuentes-Guridi \textit{et al.\ }~\cite{VIBP} generalized the original Berry
model of a spin-1/2 particle in a time varying magnetic field to a full
quantum counterpart, wherein the classical driving field was replaced by a
quantized field. A photon-dependent BP is then found in the usual
Jaynes-Cummings (JC) model. Interestingly, such a BP still exists even when
the field is at vacuum. The existence of the BP~\cite{LiuYu} in the
two-level atom system has been successfully extended to various models,
including multi-atom Dicke model~\cite{L171,L172,L173} and a multilevel atom
in a quantized field~\cite{L21}.

Surprisingly in 2012 Larson claimed that, the BP in quantum optical system
is just a result of the rotating wave approximation (RWA) and should vanish
beyond this approximation~\cite{12prl}. Note that this argument is obtained
based on a semiclassical Rabi model, wherein the field operator is simply
replaced by a complex C-number, and thus the relevant interpretation is
practically not related to the quantized light field. Alternatively we
present, in this letter, a universal formulation of the BPs for the Rabi
model and show that the non-zero BP always exists, no matter whether the RWA
is applied.

The general description of the BP in a Rabi model is given in Sec.II. Then
the validity of such a BP is verified by investigating the semiclassical
counterpart of Rabi model in Sec.~III. In Sec. IV we generalize our results
to a three-level atom in the quantized light field beyond the RWA. Finally
we present the conclusion and discussion in Sec.V.

\section{General formulation of BP in Rabi model}

The interaction of a two-level atom with a single-mode quantized field can
be generally described by the Hamiltonian

\begin{equation}
H_{0}=\omega a^{\dagger }a+\frac{\nu }{2}\sigma _{z}+\lambda (\sigma
_{+}a+\sigma _{-}a^{\dagger })+\lambda _{NR}(\sigma _{+}a^{\dagger }+\sigma
_{-}a)\text{,}
\end{equation}%
in which $\sigma _{z}$ and $\sigma _{\pm }=(\sigma _{x}+i\sigma _{y})/2$ are
called the pseudo-spin operators for the two-level atom of the
eigenfrequency $\nu $. $a^{\dagger }(a)$ denotes bosonic creation
(annihilation) operator of the single-mode quantized field with the
frequency $\omega $. Obviously when $\lambda _{NR}=0$ ,namely under the
usual RWA, the Hamiltonian reduces to that of the JC model. Correspondingly,
when $\lambda _{NR}=\lambda $ it becomes the standard Rabi-model Hamiltonian.

Following Fuentes-Guridi \textit{et al.}~\cite{VIBP} the BP can be obtained
\ in terms of an unitary transformation~\cite{VIBP}
\begin{equation}
U(\varphi )=\exp (-i\varphi a^{\dagger }a)  \label{Uni}
\end{equation}%
applied to the Hamiltonian $H_{0}$ such that

\begin{eqnarray}
H(\varphi ) &=&U(\varphi )H_{0}U^{\dagger }(\varphi )  \notag \\
&=&\omega a^{^{\prime }\dagger }a^{\prime }+\frac{\nu }{2}\sigma
_{z}+\lambda (\sigma _{+}a^{\prime }+\sigma _{-}a^{^{\prime }\dagger })
\notag \\
&&+\lambda _{NR}(\sigma _{+}a^{^{\prime }\dagger }+\sigma _{-}a^{\prime })%
\text{,}  \label{Hphii}
\end{eqnarray}%
with $a^{\prime }(\varphi )=a\exp \left( i\varphi \right) $. The eigenstates
$\left\vert \psi _{n}(\varphi )\right\rangle $ of the Hamiltonian $H(\varphi
)$ are obtained as $\left\vert \psi _{n}(\varphi )\right\rangle =U(\varphi
)\left\vert \psi _{n}\right\rangle $ with $\left\vert \psi _{n}\right\rangle
$ being the eigenstates of $H_{0}$. When the angle variable $\varphi $
slowly varies from $0$ to $2\pi $ a BP given by

\begin{equation}
\gamma _{n}=i\oint\nolimits_{c}d\varphi \left\langle \psi _{n}(\varphi
)\right\vert \frac{d}{d\varphi }\left\vert \psi _{n}(\varphi )\right\rangle
=2\pi \left\langle \psi _{n}\right\vert a^{+}a\left\vert \psi
_{n}\right\rangle  \label{Berry}
\end{equation}%
is generated for the eigenstates $\left\vert \psi _{n}(\varphi
)\right\rangle $. When $\lambda _{NR}=0$, $H_{0}$ reduces to the JC model
Hamiltonian with the ground state denoted by $\left\vert \psi
_{0}\right\rangle =\left\vert 0\right\rangle \otimes \left\vert
g\right\rangle $. Here, $\left\vert n\right\rangle $ is the Fock state of $n$
photons for the field, and $\left\vert g\right\rangle $ ($\left\vert
e\right\rangle $) is the atomic ground (excited) state. It is seen from Eq.~(%
\ref{Berry}) that the BP is zero for the ground state $\left\vert \psi
_{0}\right\rangle $, but non-zero for any excited state even if the filed is
at the vacuum. One can easily prove that the non-zero BP is practically the
half of the solid angle subtended\textbf{\ }by the traversed loop of the
eigenstate $\left\vert \psi _{n}(\varphi )\right\rangle $ on the relevant
Bloch sphere in the basis of $\left\vert n+1\right\rangle \otimes \left\vert
g\right\rangle $ and $\left\vert n\right\rangle \otimes \left\vert
e\right\rangle $. Here, the half solid angle is $\pi (1-\cos \theta _{n})$,
with $\theta _{n}$ being the angle between the eigenstate vector and the
north axis. Note that $\theta _{n}$ is also associated with the coefficient
of the eigenstate, and thus the above solid angle can be expressed as $2\pi
\lbrack \left\langle \psi _{n}\right\vert a^{+}a\left\vert \psi
_{n}\right\rangle -n]$. On the other hand, when $\lambda _{NR}=\lambda $
i.e. the case of Rabi model, any eigenstate can no longer be written as the
form of $\left\vert 0\right\rangle \otimes \left( C_{g}\left\vert
g\right\rangle +C_{e}\left\vert e\right\rangle \right) $ due to the
existence of the counter rotating wave terms. This implies that, the average
photon number of any eigenstate of the Rabi model should not be zero. As a
consequence, the BP of any eigenstate induced by the unitary transformation $%
U(\varphi )$ is always non-zero according to the Eq.~(\ref{Berry}). In fact
the non-zero BPs have been found for various eigenstates in the Rabi model~%
\cite{LiuTao}.

\section{Apparent controversy on BP in the semiclassical theory of Rabi
model and its resolution}

In this section we first of all briefly analyze how the claim of vanishing
BP~\cite{12prl} comes out for the Rabi model, and then show that it is
actually non-zero if the proper semiclassical approximation is made.

\subsection{Vanishing BPs as a result of improper semiclassical approximation%
}

Following Ref.~\cite{12prl}, we begin with the "semiclassical approximation"
of the Hamiltonian Eq. (\ref{Hphii}) (quotation mark here means improper)
i.e.,
\begin{eqnarray}
H_{C}(\varphi ) &=&\omega \left\vert \alpha \right\vert ^{2}+\frac{\nu }{2}%
\sigma _{z}  \notag \\
&&+\lambda (\alpha e^{i\varphi }\sigma _{+}+\alpha ^{\ast }e^{-i\varphi
}\sigma _{-})  \notag \\
&&+\lambda _{NR}(\alpha ^{\ast }e^{-i\varphi }\sigma _{+}+\alpha e^{i\varphi
}\sigma _{-})\text{,}  \label{HCphii}
\end{eqnarray}%
which is obtained by directly replacing the bosonic operators $a$ and $a^{+}$
with the complex C-numbers $\alpha $, $\alpha ^{\ast }$ respectively. It can
be further written as
\begin{eqnarray}
H_{C}(\varphi ) &=&\omega \left\vert \alpha \right\vert ^{2}+\left\vert
\alpha \right\vert \cos \phi (\lambda +\lambda _{NR})\sigma _{x}  \notag \\
&&+\left\vert \alpha \right\vert \sin \phi (\lambda _{NR}-\lambda )\sigma
_{y}+\frac{\nu }{2}\sigma _{z}\text{,}  \label{HCphiss}
\end{eqnarray}%
with $\alpha =\left\vert \alpha \right\vert \exp (i\varphi ^{\prime })$ and $%
\phi =\varphi +\varphi ^{\prime }$. Obviously this Hamiltonian is formally
equivalent to that of a spin-1/2 particle driven by a magnetic field: $\vec{B%
}=\left( \left\vert \alpha \right\vert \cos \phi (\lambda +\lambda _{NR})%
\text{,\quad }\left\vert \alpha \right\vert \sin \phi (\lambda _{NR}-\lambda
)\text{,\quad }\nu /2\right) $. The eigenstates of the effective Hamiltonian
can be found as a spin coherent states $\left\vert \pm \overrightarrow{n}%
\right\rangle $, where $\vec{n}\cdot \vec{\sigma}\left\vert \pm \overset{%
\rightarrow }{n}\right\rangle =\pm \left\vert \pm \overset{\rightarrow }{n}%
\right\rangle $ with $\vec{n}$ being the unit vector along the $\vec{B}$%
-direction. Therefore the slow variation of the parameter $\varphi $ from $0$
to $2\pi $ corresponds to the eigenstates $\left\vert \pm \overset{%
\rightarrow }{n}\right\rangle $\textbf{\ }traversing a loop on\textbf{\ }the
Bloch sphere and then the eigenstates will acquire BPs given by $\gamma
_{\pm }=\pm \Omega /2$, where $\Omega $ is the enclosed solid angle by the
loop. On the other hand if the eigenstates $\left\vert \pm \overset{%
\rightarrow }{n}\right\rangle $ traverse just an arc but not a closed loop,
the relevant geometric phase is zero~\cite{book,Liang}.

Specifically when $\lambda _{NR}=0$ the Hamiltonian $H_{C}(\varphi )$
becomes the "semiclassical approximation" of the JC model. Its eigenstates
are found as
\begin{equation}
\left\vert L_{+}\right\rangle =\binom{\cos \frac{\theta }{2}}{\sin \frac{%
\theta }{2}e^{-i\phi }}\text{,\qquad }\left\vert L_{-}\right\rangle =\binom{%
-\sin \frac{\theta }{2}}{\cos \frac{\theta }{2}e^{-i\phi }}\text{,}
\end{equation}%
where $\tan \theta =2\left\vert \alpha \right\vert \lambda /\Delta $, $%
\Delta =\nu -\omega $. It is seen that, when $\varphi $ varies slowly from $%
0 $ to $2\pi $ the eigenstates $\left\vert L_{\pm }\right\rangle $ acquire
the BPs $\gamma _{\pm }=\pm \Omega /2$ with the solid angle $\Omega =2\pi
(1-\cos \theta )$. While for the "semiclassical approximation" of the Rabi
model when $\lambda _{NR}=\lambda $, the ground state becomes
\begin{equation}
\left\vert R_{-}\right\rangle =\frac{1}{\sqrt{E_{-}(2E_{-}-\nu )}}\binom{%
2\left\vert \alpha \right\vert \cos \phi }{E_{-}-\nu /2},
\end{equation}%
with $E_{-}=\omega \left\vert \alpha \right\vert ^{2}-\sqrt{\nu
^{2}/4+4\left\vert \alpha \right\vert ^{2}\cos ^{2}\phi }$. Since the
elements $2\left\vert \alpha \right\vert \cos \phi $ and $E_{-}-\nu /2\neq 0$
are both real, the ground state $\left\vert R_{-}\right\rangle $ traverses
only an arc giving the zero BP when $\varphi $ varies from $0$ to $2\pi $.
Thus the conclusion of vanishing BP in the Rabi model \cite{12prl} is
recovered and it seems that the non-zero BP in the JC model is just a result
of the RWA.

However, this apparent controversy comes entirely from the improper
semiclassical approximation used in Ref.~\cite{12prl}. The Hamiltonian $%
H_{C}(\varphi )$ by directly replacing the creation (annihilation) operator $%
a^{\dagger }$\thinspace \thinspace ($a$) with a complex C-numbers $\alpha
^{\ast }$\thinspace \thinspace ($\alpha $) does not correspond to the
original one $H(\varphi )$ obtained with the unitary transformation. The
correct way to achieve the Hamiltonian $H_{C}(\varphi )$ is by the sub-space
average of $H(\varphi )$ such that $H_{C}(\varphi )=\left\langle \alpha
\right\vert H(\varphi )\left\vert \alpha \right\rangle $, where $|\alpha
\rangle $ is the optical coherent state with the usual definition $a|\alpha
\rangle =\alpha |\alpha \rangle $. Then one is able to obtain the
semiclassical ground state of $H_{C}(\varphi )$ with the standard
variational method, in which the coherent state $|\alpha \rangle $ acts as a
trial wave function. Once doing so the BP emerges in both the Rabi and JC
models independent of the RWA as it should be.

\subsection{Variational ground-state of the Rabi model and nonvanishing BP}

The proper semiclassical approximation begins with the average of the
original Rabi-mode Hamiltonian $H_{0}$ (for $\lambda _{NR}=\lambda $) in the
coherent state~\cite{Lianjl} $|\alpha \rangle $, which leads to the
following effective spin Hamiltonian
\begin{eqnarray}
H_{e}(\alpha ) &=&\left\langle \alpha \right\vert H_{0}\left\vert \alpha
\right\rangle  \notag \\
&=&\omega \left\vert \alpha \right\vert ^{2}+\frac{\nu }{2}\sigma
_{z}+\lambda (\alpha +\alpha ^{\ast })\sigma _{x}\text{.}
\end{eqnarray}
Its eigenvalues can be obtained by solving the energy eigenvalue equation,
\begin{equation}
H_{e}(\alpha )\left\vert \psi \right\rangle =E\left( \alpha \right)
\left\vert \psi \right\rangle \text{.}
\end{equation}%
The average energy
\begin{eqnarray}
E_{\pm }\left( \alpha \right) &=&\left\langle \psi _{\pm }\right\vert
H_{e}(\alpha )\left\vert \psi _{\pm }\right\rangle  \notag \\
&=&\omega \left\vert \alpha \right\vert ^{2}\pm \sqrt{\frac{\nu ^{2}}{4}%
+\lambda ^{2}\left( \alpha +\alpha ^{\ast }\right) ^{2}}
\end{eqnarray}%
is a function of $\alpha $ ,which is considered as a variational parameter
to be determined by the variation procedure, here $|\psi _{\pm }\rangle $
denotes the two eigenstates of the effective spin Hamiltonian $H_{e}(\alpha
) $. Consequently the variational ground-state energy can be obtained by
minimizing the energy function $E_{\pm }\left( \alpha \right) $. It is easy
to find that only $E_{-}(\alpha )$ leads to the true ground state and the
extremum condition is
\begin{equation}
\frac{\partial E_{-}(\alpha )}{\partial \alpha }=0\text{,}
\end{equation}%
which gives rise to the average photon number of the semiclassical
ground-state
\begin{equation}
\alpha _{gs}=\left\{
\begin{array}{c}
0\text{,} \\
\\
\sqrt{\lambda ^{2}/\omega ^{2}-\nu ^{2}/16\lambda ^{2}}\text{,}%
\end{array}%
\begin{array}{c}
\lambda \leq \lambda _{c} \\
\\
\lambda >\lambda _{c}%
\end{array}%
\right.
\end{equation}%
and the average energy
\begin{equation}
E_{gs}=\left\{
\begin{array}{c}
-\frac{\nu }{2}\text{,} \\
\\
-\frac{\lambda ^{2}}{\omega }-\frac{\omega \nu ^{2}}{16\lambda ^{2}}\text{,}%
\end{array}%
\begin{array}{c}
\lambda \leq \lambda _{c} \\
\\
\lambda >\lambda _{c}%
\end{array}%
\right.
\end{equation}%
where%
\begin{equation*}
\lambda _{c}=\frac{1}{2}\sqrt{\omega \nu }
\end{equation*}%
is the well known critical value of the quantum phase transition from the
normal to the superradiant phases in the $N$-atom Dicke model \cite%
{L172,Lianjl}. Therefore the desired semiclassical ground-state reads
\begin{eqnarray}
\left\vert \chi \right\rangle &=&\left\vert \alpha _{gs}\right\rangle
\left\vert \psi _{-}\right\rangle  \notag \\
&=&\left\{
\begin{array}{c}
\left\vert 0\right\rangle \left\vert g\right\rangle , \\
\\
\left\vert \alpha _{gs}\right\rangle \left( C_{-}\left\vert g\right\rangle
+C_{+}\left\vert e\right\rangle \right) ,%
\end{array}%
\begin{array}{c}
\lambda \leq \lambda _{c} \\
\\
\lambda >\lambda _{c}%
\end{array}%
\right.
\end{eqnarray}%
with $C_{\pm }=\sqrt{2\lambda ^{2}\mp \omega \nu /2}/2\lambda $, and the
relevant BP is calculated as
\begin{equation}
\gamma _{gs}=\left\{
\begin{array}{c}
0\text{,} \\
\\
2\pi \left\vert \alpha _{gs}\right\vert ^{2}\text{,}%
\end{array}%
\begin{array}{c}
\lambda \leq \lambda _{c} \\
\\
\lambda >\lambda _{c}%
\end{array}%
\right. \text{.}
\end{equation}%
Obviously the BP of the semiclassical ground-state in Rabi model is non-zero
when $\lambda >\lambda _{c}$. For a JC model the results are the same except
a possible shift of the critical point $\lambda _{c}$. The variational
result can be verified in a more general sense by numerical diagonalization
of the Hamiltonian $H_{0}$ with $\lambda _{NR}=\lambda $. The numerical
values of BP as a function of the dimensionless coupling constant $g=\lambda
/\omega $ are plotted in Fig.1. It may be worthwhile to remark that the
average photon-number $\alpha _{gs}$ obtained from the variational treatment
with the coherent state $|\alpha \rangle $ may have a certain amount of
deviation with respect to the accurate results of full quantum mechanical
formulation \cite{PRL105,PRL107}, and also to the numerical diagonalization
shown in Fig.1. After all, the variational result in the coherent state is
just a semiclassical approximation.

\begin{figure}[tbph]
\centering
\includegraphics[width=8cm]{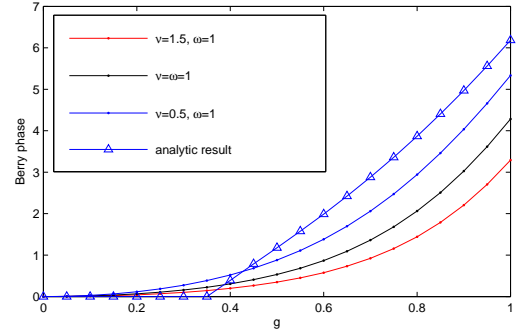}\newline
\caption{BP of the ground state in the Rabi model versus the dimensionless
coupling parameter $g=\protect\lambda /\protect\omega $ for different atomic
eigenfrequency $\protect\nu $ and field frequency $\protect\omega $.}
\label{Fig1}
\end{figure}

\section{ Ground-state BP of a three-level atom in the quantized light-field}

We now generalize the above BP formulation of two-level atom to a
three-level atom in the quantized cavity field. Without loss of the
generality, we consider the $\Lambda $-type atom with the energy eigenvalues
$E_{i}=\hbar \omega _{i},$ for $i=1,2,3$. The allowed dipole-transitions are
assumed to be $|3\rangle \leftrightarrow |2\rangle $ and $|3\rangle
\leftrightarrow |1\rangle $ but not for $|1\rangle \leftrightarrow |2\rangle
$~\cite{L21}. Beyond the RWA, the Hamiltonian of this system reads
\begin{equation}
H^{\Lambda }=H_{0}^{\Lambda }+\eta (b+b^{\dagger })(\left\vert
1\right\rangle \left\langle 3\right\vert +\left\vert 3\right\rangle
\left\langle 1\right\vert +\left\vert 2\right\rangle \left\langle
3\right\vert +\left\vert 3\right\rangle \left\langle 2\right\vert ).
\label{trie}
\end{equation}%
Here,
\begin{equation*}
H_{0}^{\Lambda }=\omega _{0}b^{\dagger }b+\omega _{1}\left\vert
1\right\rangle \left\langle 1\right\vert +\omega _{2}\left\vert
2\right\rangle \left\langle 2\right\vert +\omega _{3}\left\vert
3\right\rangle \left\langle 3\right\vert ,
\end{equation*}%
and $\eta $ is the coupling constant between atom and the field of frequency
$\omega _{0}$. $b$ and $b^{\dagger }$ denote the corresponding field
operators.

Similarly, the average of Hamiltonian $H^{\Lambda }$ in the field
coherent-state $\left\vert \beta \right\rangle $ ($b\left\vert \beta
\right\rangle =\beta \left\vert \beta \right\rangle $) becomes an effective
Hamiltonian of atomic operator only
\begin{eqnarray}
H_{e}^{\Lambda } &=&\left\langle \beta \right\vert H^{\Lambda }\left\vert
\beta \right\rangle   \notag \\
&=&\omega _{0}\left\vert \beta \right\vert ^{2}+\sum_{l}\omega
_{l}\left\vert l\right\rangle \left\langle l\right\vert +2u\eta (\left\vert
1\right\rangle \left\langle 3\right\vert +  \notag \\
&&\left\vert 3\right\rangle \left\langle 1\right\vert +\left\vert
2\right\rangle \left\langle 3\right\vert +\left\vert 3\right\rangle
\left\langle 2\right\vert )\text{,}
\end{eqnarray}%
with the complex number $\beta =u+iv$ considered as a variational parameter
to be determined. For simplicity, let us assume $\omega _{1}=\omega _{2}$,
and then the energy eigenvalue equation
\begin{equation}
H_{e}^{\Lambda }\left\vert \psi ^{\Lambda }(\beta )\right\rangle =E^{\Lambda
}(\beta )\left\vert \psi ^{\Lambda }(\beta )\right\rangle \text{,}
\end{equation}%
can be solved with the results given by
\begin{equation*}
E_{0}^{\Lambda }(\beta )=\omega _{0}\left\vert \beta \right\vert ^{2}+\omega
_{1}\text{,}
\end{equation*}%
and
\begin{equation*}
E_{\pm }^{\Lambda }(\beta )=\omega _{0}\left\vert \beta \right\vert ^{2}+%
\frac{\omega _{1}+\omega _{3}}{2}\pm \sqrt{\frac{\left( \omega _{1}-\omega
_{3}\right) ^{2}}{4}+8\eta ^{2}u^{2}}\text{.}
\end{equation*}%
One can easily check that the energy function $E_{-}^{\Lambda }(\beta )$ is
the lowest one i.e. $E_{-}^{\Lambda }(\beta )\leq E_{0}^{\Lambda }(\beta )$,
$E_{+}^{\Lambda }(\beta )$. Therefore, the ground state energy can be
determined from the extremum condition $\partial E_{-}^{\Lambda }(\beta
)/\partial \beta =0$ and the result is
\begin{equation}
E_{gs}^{\Lambda }=\left\{
\begin{array}{c}
\omega _{1}\text{,} \\
\\
\frac{-2\eta ^{2}}{\omega _{0}}-\frac{\omega _{0}\left( \omega _{1}-\omega
_{3}\right) ^{2}}{32\eta ^{2}}+\frac{\omega _{1}+\omega _{3}}{2}\text{,}%
\end{array}%
\begin{array}{c}
\eta \leq F \\
\\
\eta >F%
\end{array}%
\right. \text{{}}
\end{equation}%
with the relevant variational parameter found as
\begin{equation}
\beta _{gs}=\left\{
\begin{array}{c}
0\text{,} \\
\\
\sqrt{\frac{2\eta ^{2}}{\omega _{0}^{2}}-\frac{\left( \omega _{1}-\omega
_{3}\right) ^{2}}{32\eta ^{2}}}\text{,}%
\end{array}%
\begin{array}{c}
\eta \leq F \\
\\
\eta >F%
\end{array}%
\right.
\end{equation}%
where $F=\sqrt{\omega _{0}\left( \omega _{1}-\omega _{3}\right) /8}$.
Consequently, we apply a unitary transformation $U(\tau )=\exp (-i\tau
b^{\dagger }b)$ to the Hamiltonian $H^{\Lambda }$ and obtain
\begin{eqnarray}
H^{\Lambda }(\tau ) &=&U(\tau )H^{\Lambda }U^{\dagger }(\tau )  \notag \\
&=&H_{0}^{\Lambda }+\eta (be^{i\tau }+b^{\dagger }e^{-i\tau })(\left\vert
1\right\rangle \left\langle 3\right\vert   \notag \\
&&+\left\vert 3\right\rangle \left\langle 1\right\vert +\left\vert
2\right\rangle \left\langle 3\right\vert +\left\vert 3\right\rangle
\left\langle 2\right\vert )\text{.}
\end{eqnarray}%
Thus, when $\tau $ varies slowly from $0$ to $2\pi $, the semiclassical
ground state $\left\vert \psi _{gs}^{\Lambda }(\tau )\right\rangle =U(\tau
)\left\vert \psi _{gs}^{\Lambda }\right\rangle $ will acquire a BP evaluated
as
\begin{eqnarray}
\gamma _{gs}^{\Lambda } &=&i\oint\nolimits_{c}d\tau \left\langle \psi
_{gs}^{\Lambda }(\tau )\right\vert \frac{d}{d\tau }\left\vert \psi
_{gs}^{\Lambda }(\tau )\right\rangle   \notag \\
&=&2\pi \left\langle \psi _{gs}^{\Lambda }\right\vert b^{+}b\left\vert \psi
_{gs}^{\Lambda }\right\rangle \text{,}
\end{eqnarray}%
which leads to the result
\begin{eqnarray}
\gamma _{gs}^{\Lambda } &=&2\pi \left\vert \beta _{gs}\right\vert ^{2}
\notag \\
&=&\left\{
\begin{array}{c}
0\text{,} \\
\\
\frac{2\eta ^{2}}{\omega _{0}^{2}}-\frac{\left( \omega _{1}-\omega
_{3}\right) ^{2}}{32\eta ^{2}}\text{,}%
\end{array}%
\begin{array}{c}
\eta \leq F \\
\\
\eta >F%
\end{array}%
\right.
\end{eqnarray}%
Again, the BP depends on the coupling constant $\eta $ as that in the
two-level case. As a comparison the BP values are also evaluated by the
numerical diagonalization of the Hamiltonian $H^{\Lambda }$. Fig.~2 shows
the plots of ground-state BPs versus the dimensionless coupling constant $%
g^{\prime }=\eta /\omega _{0}$ . It is seen that the numerical results are
qualitatively in agreement with our semiclassical analysis.
\begin{figure}[tbph]
\centering
\includegraphics[width=8cm]{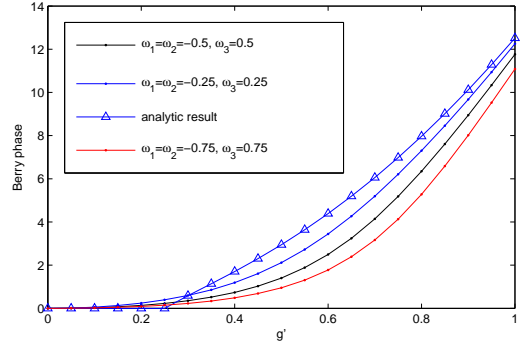}\newline
\caption{Ground-state BP in the three-level atom system versus the
dimensionless atom-field coupling constant $g^{\prime }$ for $\protect\omega %
_{1}=\protect\omega _{2}=-0.25$, $\protect\omega _{3}=0.25$.}
\label{Fig2}
\end{figure}

\section{Conclusion and Discussion}

We present in this letter a general formulation of the BPs for both the JC
model (with RWA) and the Rabi model (without the RWA). In the semiclassical
approximation the variational ground-state and the related BP for both Rabi
and JC models are obtained analytically in a consistent manner. We argue
that the vanishing BP of the Rabi model \cite{12prl} is due to the improper
semiclassical approximation with simply replacing the bosonic operator by a
complex C-number since the resulted Hamiltonian does not correspond to the
original one in the semiclassical level. The valid effective spin
Hamiltonian $H_{e}(\alpha )$ in the semiclassical approximation is achieved
by the average in the field coherent state and thus the macroscopic
(semiclassical) quantum state should be obtained by means of the standard
variational method. We show in this letter that the BP for a generic Rabi
model is indeed non-zero. This observation is also been generalized to a
three-level atom in the quantized cavity field.

The BP in the variational ground-state, which is a displaced vacuum (i.e.
coherent state), may possess a simple interpretation as suggested by one
Referee of the paper that the action of the unitary transformation Eq.~(\ref%
{Uni}) is to take this coherent state around the origin in phase space and
thus the emerging BP is just the one of a harmonic oscillator. Following the
Ref.~\cite{VIBP}, the BPs beyond the RWA, obtained in the present work,
should be also verified experimentally, in principle, with the usual cavity
quantum electrodynamic system. The analogous experiment to measure the BP
has been also designed~in a single solid-state spin-quibt~\cite{shiyan}.

\section*{Acknowledgements}

This work was supported in part by the National Natural Science Foundation
of China, under Grants No. 61301031, 11174373, U1330201, 11275118 and the
National Foundation Research Program of China, through Grant No.
2010CB923104.

\section*{References}

\bibliography{wangminghao}

\end{document}